\documentstyle[pra,aps,cite]{revtex}

\begin{document}

\title{Hydrogen molecule in a magnetic field: 
The lowest states of the $\Pi$ manifold and the global ground state of the parallel configuration}
\author{T.Detmer, P. Schmelcher and L. S. Cederbaum}
\address{Theoretische Chemie, Physikalisch--Chemisches Institut,\\
Universit\"at Heidelberg, INF 253, D--69120 Heidelberg,\\
Federal Republic of Germany}
\maketitle

\begin{abstract}
The electronic structure of the hydrogen molecule in a magnetic field is 
investigated for parallel internuclear and magnetic field axes. 
The lowest states of the $\Pi$ manifold are studied for spin singlet and triplet$\left( M_s = -1 \right) $  
as well as gerade and ungerade parity for a broad range of field strengths 
$0 \leq B \leq 100\;a.u.$ 
For both states with gerade parity we observe 
a monotonous decrease in the 
dissociation energy with increasing field strength up to $B = 0.1\;a.u.$ 
and metastable states with respect to the dissociation into two H atoms occur for 
a certain range of field strengths.  
For both states with ungerade parity we observe a strong increase in the 
dissociation energy with increasing field strength above some critical field strength $B_c$. 
As a major result we determine the transition field 
strengths for the crossings among the lowest $^1\Sigma_g$, $^3\Sigma_u$ and $^3\Pi_u$ states.  
The global ground state for $B \lesssim 0.18\;a.u.$ is the strongly bound $^1\Sigma_g$ state. 
The crossings of the $^1\Sigma_g$ with the $^3\Sigma_u$ and $^3\Pi_u$ state occur at   
$B \approx 0.18$ and $B \approx0.39\;a.u.$, respectively. 
The transition between the $^3\Sigma_u$ and 
$^3\Pi_u$ state occurs at $B \approx 12.3\;a.u.$ Therefore, the {\sl global ground state} of the 
hydrogen molecule for the parallel configuration is the {\sl unbound} $^3\Sigma_u$ state for $0.18 \lesssim B \lesssim 12.3\;a.u.$ 
The ground state for $B \gtrsim 12.3\;a.u.$ is the strongly bound $^3\Pi_u$ state. 
This result is of great relevance to the chemistry in the atmospheres of magnetic 
white dwarfs and neutron stars. 

\end{abstract}

\section{Introduction} \label{introduction}

The behavior and structure of matter in the presence of strong external magnetic fields 
is a research area of increasing interest. 
This increasing interest is motivated by the occurence of strong fields and strong field effects 
in  different branches of physics like astrophysics \cite{ostriker:1968,kemp:1970,truemper:1977},
atomic and molecular physics of Rydberg states \cite{friedrich:1989} and certain areas 
of solid state physics like excitons and/or quantum nano structures \cite{chiu:1974}. 

A number of theoretical investigations were performed concerning the properties of 
atomic and molecular systems in strong magnetic fields. Most of them however deal with 
the hydrogen atom. For molecular systems only the electronic structure of the $H_2^+$ ion 
was investigated in some detail (see Refs. 
\cite{wille:1988,kappes1:1996,kappes2:1996,kappes3:1996,kravchenko:1997} 
and references therein). Very interesting phenomena can be observed 
already for this simple diatomic system. For the ground state of the $H_2^+$ molecule 
the dissociation energy increases and the equilibrium internuclear distance simultaneously 
decreases with increasing field strength. 
Furthermore it was shown \cite{kappes1:1994,kappes:1995} that a certain class of excited 
electronic states, which possess a purely repulsive potential energy surface 
in the absence of a magnetic field, acquire a well-pronounced potential well 
in a sufficiently strong magnetic field.
Moreover the electronic potential energies depend not only on the internuclear distance 
but also on the angle between the magnetic field and molecular axes which leads to a very 
complex topological behavior of the corresponding potential energy surfaces 
\cite{kappes1:1996,kappes2:1996,kappes3:1996}. 

In contrast to the $H_2^+$ ion there exist only a few investigations dealing with  
the electronic structure of the hydrogen molecule in the presence of a strong magnetic field. 
Highly excited states of $H_2$ were studied for a field strength of $4.7\;T$ in Ref. 
\cite{monteiro:1990}. 
For intermediate field strengths two studies of almost qualitative character 
investigate the potential energy curve (PEC) of the lowest $^1\Sigma_g$ state 
\cite{basile:1987,turbiner:1983}. 
A few investigations were performed in the high field limit 
\cite{korolev:1992,lai:1992,ortiz:1995,lai:1996}, where the magnetic forces 
dominate over the Coulomb forces and therefore 
several approximations can be performed. 
Very recently a first step has been done in order to elucidate the electronic structure 
of the $H_2$ molecule for the parallel configuration, i.e. for parallel internuclear and 
magnetic field axes \cite{detmer1:1997}. In that investigation the ground states of the 
$\Sigma$ manifold were studied for gerade and ungerade parity as well as singlet and triplet states. 
Hereby accurate adiabatic electronic energies were obtained for a broad range of field 
strengths from field free space up to 
strong magnetic fields of $100\;a.u.$ 
A variety of interesting effects were revealed. As in the case of the $H_2^+$ ion,
the lowest strongly bound states, i.e. the lowest $^1\Sigma_g$ , $^3\Sigma_g$  and $^1\Sigma_u$  state, show a decrease of the 
bond length and an increase in the dissociation energy for sufficiently 
strong fields. Furthermore a change in the dissociation channel occurs for the lowest $^1\Sigma_u$  state 
between $B = 10.0$ and $20.0\;a.u.$ due to the existence of strongly bound $H^-$ states 
in the presence of a magnetic field. The $^3\Sigma_g$  state was shown to exhibit an additional 
outer minimum for intermediate field strengths which could provide vibrationally bound states. 

An important result of Ref. \cite{detmer1:1997} is the change of the ground state from the 
lowest $^1\Sigma_g$ state to the lowest $^3\Sigma_u$ state between $B = 0.1$ and $0.2\;a.u.$
This crossing is of particular relevance 
for the existence of molecular hydrogen in the vicinity of white dwarfs. 
The $^3\Sigma_u$  state is an unbound state and possesses only a very shallow 
van der Waals minimum which does not provide any vibrational level. 
Therefore, the ground state of the hydrogen molecule for the parallel configuration is an 
unbound state for 
$B \gtrsim 0.2\;a.u.$ up to some much higher critical field strength $B_c$ 
which is not known exactly. 
In Ref. \cite{ortiz:1995} it has been shown that for very strong fields 
($B \gtrsim 3\times10^3\;a.u.$) the strongly bound $^3\Pi_u$ state is the global ground state of 
the hydrogen molecule oriented parallel to the magnetic field. 
Therefore a second transition which involves the unbound $^3\Sigma_u$ 
and the strongly bound $^3\Pi_u$ state has to occur at some field strength 
between $0.2$ and $3\times10^3\;a.u.$

The above considerations show that detailed studies of the electronic 
properties of the hydrogen molecule in a magnetic field are very desirable. 
The present investigation deals with the electronic structure 
of the lowest states of the important $\Pi$ subspace, i.e. the lowest singlet 
and triplet$\left( M_s = -1\right) $ states with 
gerade and ungerade parity. 
This subspace contains, as mentioned above, the global ground state of the $H_2$ molecule 
for sufficiently strong magnetic fields. 
We hereby consider the case of parallel internuclear and 
magnetic field axes. The parallel configuration is distinct by its high symmetry compared to the case of 
an arbitrary angle between the internuclear and magnetic field axes and is expected 
to play an important role for the electronic 
structure of the hydrogen molecule \cite{schmelcher:1990}. 
The results of our calculations include accurate adiabatic PECs for the complete 
range of field strengths $0 \leq B \leq 100\;a.u.$ We present detailed data 
for the total and dissociation energies, equilibrium internuclear distances and positions of 
maxima for the corresponding electronic states. Moreover we provide a discussion of the global ground 
state of the parallel configuration and give the transition field strengths for the crossings 
between the $^1\Sigma_g$ , $^3\Sigma_u$  and $^3\Pi_u$ states. 

In detail the paper is organized as follows. In Sec. \ref{theory} we describe the theoretical aspects 
of the present investigation, including a discussion of the Hamiltonian and a description 
of the basis set of our CI calculations. In Sec. \ref{fieldfree} 
we discuss the general aspects of the potential energy curves (PECs) in field free space 
for the lowest states of the $\Pi$ subspace. Sec. \ref{bfield} contains a detailed  investigation 
of the electronic structure of the lowest $\Pi$ states in the presence of a magnetic field 
for the range $0.001 \leq B \leq 100\;a.u.$ Finally the global ground state of the 
hydrogen molecule oriented parallel to the magnetic field is studied in Sec. \ref{ground}. 
The summary and conclusions are given in Sec. \ref{summary}. 

\section{Theoretical aspects} \label{theory}

Our starting point is the total nonrelativistic molecular Hamiltonian in Cartesian coordinates.
The total pseudomomentum is a constant of motion and therefore commutes with the Hamiltonian 
\cite{johnson:1983,avron:1978}. 
For that reason the Hamiltonian can be simplified by performing a
so-called pseudoseparation of the center of mass motion 
\cite{schmelcher2:1988,schmelcher:1994,johnson:1983} which introduces 
the center of mass coordinate and the conserved 
pseudomomentum as a pair of canonical conjugated variables. 
Further simplifications can be achieved by a consecutive series of unitary transformations
\cite{schmelcher2:1988,schmelcher:1994}.

In order to separate the electronic and nuclear motion we perform the 
Born-Oppenheimer approximation in the presence of a magnetic field
\cite{schmelcher1:1988,schmelcher2:1988,schmelcher:1994}. 
As a first order approximation we assume infinitely heavy masses for the nuclei.  
The origin of our coordinate system coincides with the midpoint of 
the internuclear axis of the hydrogen molecule 
and the protons are located on the $z$ axis.  
The magnetic field is chosen parallel to the $z$ axis of our coordinate system and
the symmetric gauge is adopted for the vector potential.
The gyromagnetic factor of the electron is chosen to be equal to two.
The Hamiltonian, therefore, takes on the following appearance:
\begin{equation}
H =\sum\limits_{i=1}^2 \left\{\frac{1}{2}\bbox{p}_i^2 
+ \frac{1}{8}\left( \bbox{B}\times \bbox{r}_i\right)^2 
+ \frac{1}{2}\bbox{L}_i \bbox{B}- \frac{1}{|\bbox{r}_i - \bbox{R}/2|} 
- \frac{1}{|\bbox{r}_i + \bbox{R}/2|} \right\} \\
+ \frac{1}{|\bbox{r}_1 - \bbox{r}_2|} + \frac{1}{R} 
+ \bbox{S} \bbox{B}\label{form1}
\end{equation}
The symbols $\bbox{r}_i$, $\bbox{p}_i$ and $\bbox{L}_i$ denote the position vectors, the 
canonical conjugated momenta and the angular momenta of the two electrons, respectively.
$\bbox{B}$ and $\bbox{R}$ are the vectors of the magnetic field and internuclear distance, 
respectively and $R$ denotes the magnitude of $\bbox{R}$.
With $\bbox{S}$ we denote the vector of the total electronic spin.
Throughout the paper we will use atomic units.

The Hamiltonian (\ref{form1}) commutes with the following operators:
the parity operator $P$, 
the projection $L_z$ of the electronic angular momentum on the internuclear axis, 
the square $S^2$ of the total electronic spin and 
the projection $S_z$ of the total electronic spin on the internuclear axis. 
In the case of field free space we encounter an additional independent symmetry namely the reflections 
of the electronic coordinates at the $xz$ ($\sigma_v$) plane. 
The eigenfunctions possess the corresponding eigenvalues $\pm1$.
This symmetry does not hold in the presence of a magnetic field!
Therefore, the resulting symmetry groups for the hydrogen molecule are $D_{\infty h}$ in the case of
field free space and $C_{\infty h}$ in the presence of a magnetic field \cite{schmelcher:1990}.

In order to solve the fixed-nuclei electronic Schr\"odinger equation 
belonging to the Hamiltonian (\ref{form1}) we expand the electronic 
eigenfunctions in terms of molecular configurations.
In a first step the total electronic wave function $\Psi_{ges}$  
is written as a product of its spatial part $\Psi$ and its spin part 
$\chi$, i.e. we have $\Psi_{ges} = \Psi \chi$.
For the spatial part $\Psi$ of the wave function we use the LCAO-MO-ansatz, i.e.
we decompose $\Psi$ with respect to molecular orbital configurations $\psi$ of $H_2$, 
which respect the corresponding symmetries (see above) and the Pauli principle: 
\begin{eqnarray}
\Psi &=& \sum\limits_{i,j} c_{ij} \left[\psi_{ij}\left(\bbox{r}_1,\bbox{r}_2\right) \pm 
\psi_{ij}\left(\bbox{r}_2,\bbox{r}_1\right)\right] \nonumber \\
&=& 
\sum\limits_{i,j} c_{ij} \left[\Phi_i\left(\bbox{r_1}\right)\Phi_j\left(\bbox{r_2}\right) 
\pm \Phi_i\left(\bbox{r_2}\right)\Phi_j\left(\bbox{r_1}\right)\right] \nonumber
\end{eqnarray}
The molecular orbital configurations $\psi_{ij}$ of $H_2$ are products of the corresponding 
one-electron $H_{2}^+$ molecular orbitals $\Phi_i$ and $\Phi_j$.
The $H_{2}^+$ molecular orbitals are built from atomic orbitals 
centered at each nucleus. A key ingredient of this procedure is a
basis set of nonorthogonal optimized 
nonspherical Gaussian atomic orbitals which has been established previously 
\cite{schmelcher3:1988,kappes2:1994}. 
For the case of a $H_2-$molecule parallel to the magnetic field 
these basis functions read as follows:

\begin{equation}
\phi^m_{kl} \left(\rho,z,\alpha,\beta,\pm R/2 \right) = 
\rho^{|m|+2k} \left(z\mp R/2\right)^l exp\left\{-\alpha \rho^2 - \beta\left( z\mp R/2 \right)^2 \right\} 
exp\left\{im\phi\right\} \label{form2} 
\end{equation}

The symbols $\rho = x^2+y^2$ and $z$ denote the electronic coordinates.   
$m$, $k$ and $l$ are parameter depending on the subspace of the H-atom 
for which the basis functions have been optimized and 
$\alpha$ and $\beta$ are variational parameters. 
For a more detailed description of the construction of the molecular 
electronic wave function we refer the reader to Ref. \cite{detmer1:1997}.

In order to determine the molecular electronic wave function of $H_2$ 
we use the variational principle which means that we minimize the variational integral
$\frac{\int \Psi^* H \Psi}{\int \Psi^* \Psi}$ by varying the 
coefficients $c_i$. 
The resulting generalized eigenvalue problem reads as follows:
\begin{equation}
\left(\underline{H} - \epsilon\underline{S}\right)\bbox{c} = \bbox{0} \label{form3}
\end{equation}
where the Hamiltonian matrix $\underline{H}$ is real and symmetric and the overlap matrix
is real, symmetric and positiv definite. The vector $\bbox{c}$ contains the expansion coefficients.
The matrix elements of the Hamiltonian matrix and the overlap matrix
are certain combinations of matrix elements with respect to 
the optimized nonspherical Gaussian atomic orbitals. 
A description of the technique of the evaluation of these 
matrix elements is given in Ref. \cite{detmer1:1997}.

For the numerical solution of the eigenvalue problem (\ref{form3}) we
used the standard NAG library.
The positions, i.e. internuclear distances, of the maxima and 
the minima in the PECs were determined with an accuracy 
of $10^{-2} a.u.$ Herefore about 350 points were calculated on an average for each PEC. 
The typical dimension of the Hamiltonian matrix for each $\Pi$ subspace varies between
approximately 1700 and 3300 depending on the magnetic field strength.  
The overall accuracy of our results in the energy is estimated to be typically 
better than $10^{-4}$ and for some cases better than $10^{-4}$.
Depending on the dimension of the Hamiltonian matrix, is takes between 70 and 250 minutes for
simultaneously calculating one point of a PEC of each $\Pi-$ subspace on a IBM RS6000 computer. 

\section{The ground states of the $\Pi-$manifold in field free space} \label{fieldfree}

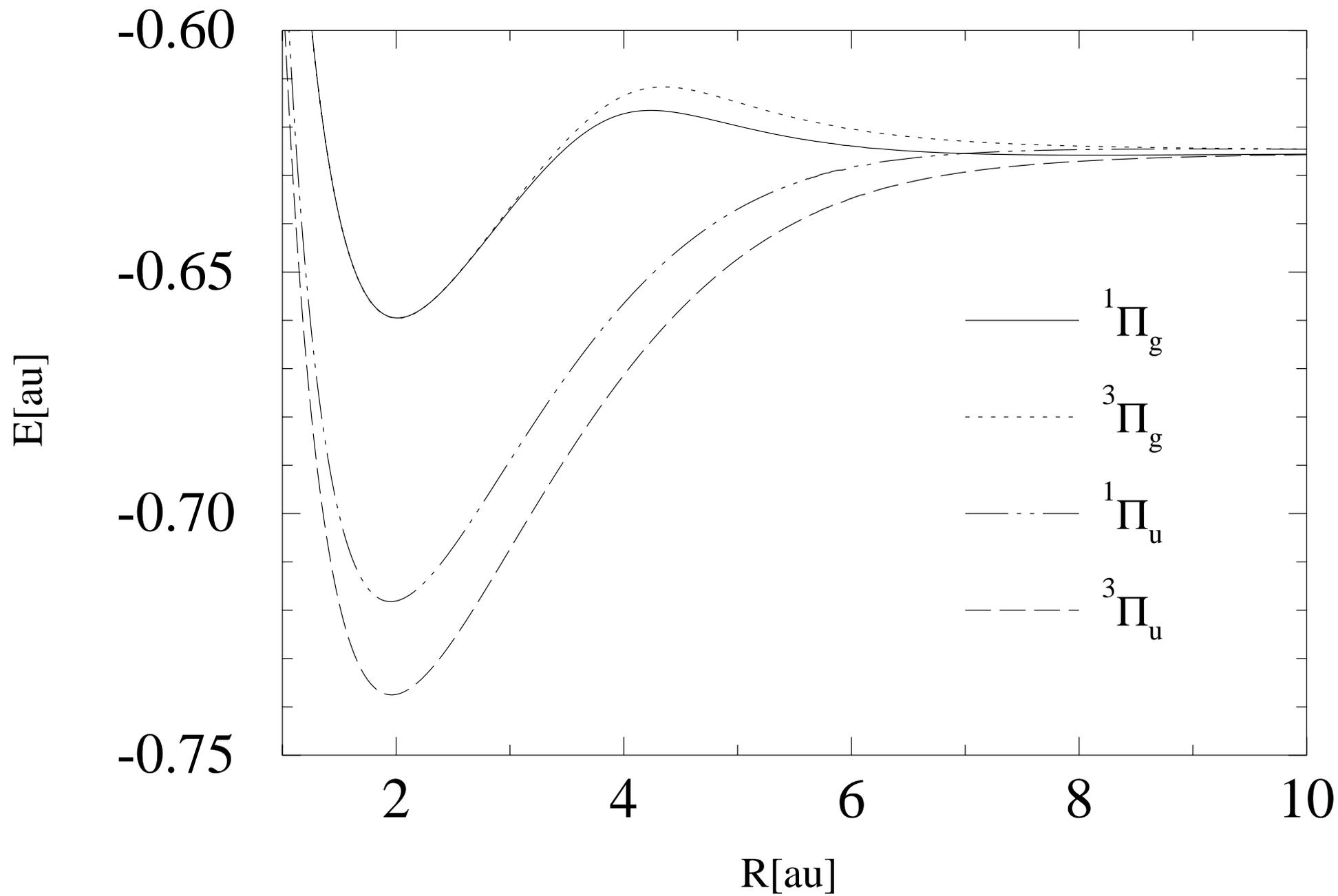
\begin{figure}
\caption{PECs (total energy) for the lowest $^1\Pi_g$,
 $^3\Pi_g$, $^1\Pi_u$ and $^3\Pi_u$ states in field free space}
\label{fig1}
\end{figure}

Let us start with the discussion of the general properties 
of the low lying $\Pi$ states in field free space whose PECs are given in Fig. \ref{fig1}. 
A detailed description 
of these states can be found in Ref. \cite{mulliken:1966}. 
At large internuclear distances both the lowest 
$^1\Pi_g$ as well as the $^3\Pi_g$ state can be approximately described by a 
Heitler-London type wave function consisting of the orbital configuration 
$\left(1s2p\pi\right)_g$. An avoided crossing with the attractive $\left( 1\sigma_g3d\pi \right) $ type state 
leads to the occurence of a minimum in both of the $\Pi_g$ PECs (see Fig. \ref{fig1}). 
For the singlet $^1\Pi_g$ state we encounter a second minimum of van der Waals type 
at large internuclear distances which has been shown to accomodate several 
vibrational levels within the Born-Oppenheimer approximation \cite{kolos:1977}. 
The united atom limit for the $^1\Pi_g$ and $^3\Pi_g$ state is the 
$^1D\;1s3d$ and $^3D\;1s3d$ helium state, respectively. 
A remarkable feature of the $^1\Pi_g$ and $^3\Pi_g$ state is the fact, that their 
wave functions are almost indistinguishable for small internuclear distances, i.e. 
$R \lesssim 2.5\;a.u.$ (see also their PECs in Fig. \ref{fig1}). 
For $R \lesssim 2.3\;a.u.$ the triplet state has a slightly lower energy than the singlet. 
The energy difference between the singlet and triplet state at the 
equilibrium internuclear distance amounts to $4.92\times10^{-5}\;a.u.$ 
and reaches a maximum of $6.38\times10^{-5}\;a.u.$ at 
$R \approx 1.6\;a.u.$ With further decreasing internuclear distance 
the energy difference decreases finally reaching a value of 
$1.55\times10^{-5}\;a.u.$ in the united atom limit \cite{kolos:1977}. 

The $\Pi_u$ states whose PECs are also illustrated in Fig. \ref{fig1} can 
be represented by a Heitler-London type wave function 
$\left(1s2p\pi\right)_u$ for large internuclear distances. 
Near the equilibrium internuclear distance the wave function can approximately be described 
by a $\left( 1\sigma_g2p \pi \right) $ type configuration. 
The PEC of the $^1\Pi_u$ state exhibits a hump at $R \approx 9\;a.u.$ due to 
a first order London dispersion force. 
For the $^1\Pi_u$ and $^3\Pi_u$ state the united atom limit is given by the 
$^1P\;1s2p$ and $^3P\;1s2p$ helium state, respectively. 
An overall feature of the lowest $\Pi$ states with ungerade parity is the fact that the PEC 
of the $^3\Pi_u$ triplet state is lower in energy than the corresponding PEC of the 
$^1\Pi_u$ singlet state. 
Extensive studies have been performed in order to explain the difference in 
the energy of the $^1\Pi_u$ and $^3\Pi_u$ states, i.e. 
the singlet-triplet energy splitting. Intuitively one would expect 
the interelectronic repulsion being smaller for the triplet state than for the singlet 
due to the zero of the wave function if the coordinates of both electrons are equal. 
However, it has been shown that, similar to the case of the $^1P$ and $^3P$ states of helium, 
the interelectronic repulsion is larger in the triplet state. 
The greater stability, i.e. lower energy, of the triplet state occurs due to a 
larger electron-nucleus attraction energy \cite{colbourn:1973}. 
With increasing internuclear distances both the PECs of the $^1\Pi_g$ and $^3\Pi_u$ state as well as 
those of the $^1\Pi_u$ and $^3\Pi_g$ state approach each other and finally end up in 
the same separated atom limit which is $H\left(1s\right) + H\left(2p\right)$.  

After describing the general aspects of the PECs for the lowest states of the 
four $\Pi$ subspaces let us compare our numerical results for these states with the existing data in the 
literature. For the $^1\Pi_g$ state very accurate energies within the 
Born-Oppenheimer approximation were obtained by Wolniewicz \cite{wolniewicz2:1995} 
and Kolos and Rychlewski \cite{kolos:1977}. 
For the equilibrium internuclear distance $R = 2.01\;a.u.$ we obtained a total energy of 
$-0.69501\;a.u.$ which yields a dissociation energy of $0.034502\;a.u.$ (cf. Table \ref{tab1}). 
This corresponds to a relative accuracy 
in the total energy of $3.2\times10^{-5}$ 
compared to the result in \cite{wolniewicz2:1995}. 
The relative accuracy of our data further improves with increasing internuclear distance. 
As an example we mention $R = 12\;a.u.$ where the relative accuracy is $2.4\times10^{-6}$. 
For the second minimum we obtained an equilibrium internuclear distance 
of $8.14\;a.u.$ with a total energy of $-0.625803\;a.u.$ The dissociation energy 
for this minimum therefore amounts to $8.038\times10^{-4}\;a.u.$ 
The maximum in the PEC with a total energy of $-0.616547\;a.u.$ 
occurs at an internuclear distance of $4.24\;a.u.$

As a reference for the Born-Oppenheimer energies of the $^1\Pi_u$ state we use the 
data given in Ref. \cite{wolniewicz3:1995}. 
Results of our calculations concerning the PEC of the $^1\Pi_u$ state 
are presented in Table \ref{tab3}. 
At the equilibrium internuclear distance $R_{eq} = 1.95\;a.u.$ our calculations 
yield a total energy of $-0.718219\;a.u.$ corresponding to a dissociation energy of 
$0.093220\;a.u.$ The relative accuracy compared to the data given in 
Ref. \cite{wolniewicz3:1995} is $2.4\times10^{-4}$ at $R_{eq}$ and $2.7\times10^{-6}$ 
at $R = 12\;a.u.$ As in the case of the $^1\Pi_g$ state the acccuracy of our calculations increases  
with increasing internuclear distances. The position of the maximum has been 
determined to be $R_{max} = 9.03\;a.u.$ and the total energy at $R_{max}$ evaluates to  $-0.624528\;a.u.$ 

Accurate Born-Oppenheimer energies for the lowest triplet states can be found in Ref. \cite{kolos:1977}. 
Data of the present investigation for the total energies and 
positions of maxima and minima are given in Table \ref{tab2} 
for the $^3\Pi_g$ and Table \ref{tab4} for the $^3\Pi_u$ state, respectively. 
Within our calculations the total energy of the $^3\Pi_g$ state at the 
equilibrium internuclear distance $R = 2.01\;a.u.$ amounts to $-0.659553\;a.u.$ 
Therefore, the relative accuracy of the total energy compared to the data given in Ref. \cite{kolos:1977} 
is $2.2\times10^{-5}$. For the $^3\Pi_u$ state an equilibrium internuclear distance of 
$R = 1.96\;a.u.$ with a total energy of $-0.737521\;a.u.$ has been obtained corresponding 
to a relative error of $6.3\times10^{-5}$. For both states the accuracy for larger 
internuclear distances is of the order of magnitude of $10^{-6}$. 

\section{The lowest $\Pi-$states in the presence of a magnetic field} \label{bfield}
 
\subsection{The $^1\Pi_g$ state}

First of all we consider the dissociation channel of the $^1\Pi_g$ state  in the presence of a 
magnetic field. In our notation atomic hydrogen states are labeled by $H\left(m_a^{\pi_a}\right)$, 
where $m_a $ is the atomic magnetic quantum number and $\pi_a$ the atomic $z$ parity. 
For the entire range of field strengths $0 < B \leq 100\;a.u.$ 
the dissociation channel of the $^1\Pi_g$ state  is given by 
$H_2 \rightarrow H\left(0^+\right) + H\left(1^+\right)$. 
This means that the energy in the dissociation limit corresponds to the total energy 
of two hydrogen atoms in the lowest electronic state within the $\left(0^+\right)$ 
and $\left(1^+\right)$ subspace, respectively. This dissociation channel holds also 
for all the other electronic $\Pi$ states considered in the present work. 

Let us now investigate the PEC of the $^1\Pi_g$ state  with varying field strength 
which is illustrated in Fig. 2a. 
For the first minimum/well we observe that the corresponding depth 
decreases strongly with increasing  magnetic field strength. 
The dissociation energy, which amounts to $E_{d1} = 0.034502\;a.u.$ in field free space, 
monotonously decreases to $E_{d1} = 0.014460\;a.u.$ for $B= 0.1\;a.u.$ 
The corresponding equilibrium internuclear distance $R_{eq1}$ hereby 
remains approximately constant up to a field 
strength of $B \approx 0.01\;a.u.$ With further increasing field strength we observe a minor 
increase to $R_{eq1} = 2.04\;a.u.$ for $B = 0.1\;a.u.$ 
Between $B = 0.1$ and $0.5\;a.u.$ we observe a drastical change in the shape of the PEC. 
For a field strength of $0.2\;a.u.$ 
we encounter a metastable state with respect to the dissociation into two hydrogen atoms. 
The minimum in the corresponding PEC occurs at an internuclear distance 
$R_{eq1} = 2.09\;a.u.$ and the difference between 
the total energies at $R_{eq1}$ and $R_{max}$ evaluates to $E_{t1} - E_{max} = 0.011130\;a.u.$ 
For even larger field strengths, i.e. $B \gtrsim 0.5\;a.u.$, the first minimum disappears.  
Simultaneously we observe a moderate change in the position of the maximum 
$R_{max}$ in the regime $0 \leq B \lesssim 0.2\;a.u.$ With increasing magnetic field strength 
$R_{max}$ is shifted to decreasingly smaller internuclear distances, i.e. from 
$R_{max} = 4.24\;a.u.$ in field free space to $R_{max} = 3.28\;a.u.$ for a field strength of $0.2\;a.u.$  
Fig. 2a shows the PECs of the $^1\Pi_g$ state for different field strengths 
illustrating the shape of the PEC near the first equilibrium internuclear distance. 
The corresponding data concerning the positions of the maxima and minima, total and 
dissociation energies, and total energies in the separated atom limit are presented in Table \ref{tab1}.

Despite the fact that the first minimum and 
the maximum of the PEC disappear for field strengths $B \gtrsim 0.5\;a.u.$ 
the PECs possess some interesting features for larger field strengths. 
In the following discussion we focus on the rough shape of the PECs shown in Fig. 2a 
for $B \gtrsim 0.5\;a.u.$ 
From Fig. 2a we observe that the PEC of the $^1\Pi_g$ state 
exhibits two turning points at $R \approx 2.3\;a.u.$ and $R \approx 3.2\;a.u.$ 
for the field strength $B = 0.5\;a.u.$ 
For $B = 1\;a.u.$ these two turning points are shifted to 
smaller internuclear distances, e.g. $R \approx 2.2\;a.u.$ and $R \approx 2.4\;a.u.$ 
From Fig. 2a we see that these turning points 
are much less pronounced for a field strength of $B = 1$ than for $0.5\;a.u.$ which is also  
confirmed by an investigation of the corresponding first derivatives of the above PECs. 
With further increasing field strength, i.e. in the range $2 \lesssim B \lesssim 20\;a.u.$, 
no turning points can be found in the PEC of the $^1\Pi_g$ state. Only for field strengths 
$B \gtrsim 50\;a.u.$ two turning points again exists 
which however cannot be seen in the PEC for $B = 100\;a.u.$ in Fig. 2a since they occur 
for very small values of the internuclear distance. This behavior is of importance 
for the singlet-triplet energy splitting between the PECs of the $^1\Pi_g$ and 
$^3\Pi_g\;(M_s = 0)$ state which will be discussed in the following subsection. 

In addition to the deep minimum at $R_{eq1} = 2.01\;a.u.$ in the 
PEC of the $^1\Pi_g$ state, a second minimum 
exists which in field free space is very shallow and located at $R_{eq2} = 8.14\;a.u.$ 
The corresponding region of the PECs is illustrated in Fig. 2b. 
In the presence of a magnetic field this minimum becomes more and more pronounced 
with increasing field strength.  The depth of the well monotonously increases 
up to a field strength of $B = 1\;a.u.$ (c.f. Table \ref{tab1}). 
The dissociation energy $E_{d2}$ for $B = 1\;a.u.$ amounts to $2.853\times10^{-3}$ 
which is more than three times as 
much as the corresponding dissociation energy in field free space. 
For field strengths $B \gtrsim 2\;a.u.$ the dissociation energy decreases monotonously. 
The corresponding equilibrium internuclear distance $R_{eq2}$ 
for this second minimum first decreases with increasing 
field strength from $8.14\;a.u.$ in field free 
space to $4.05\;a.u.$ at $B = 10\;a.u.$ but exhibits a minor increase for $B\gtrsim20\;a.u.$ 
with further increasing magnetic field strength. 
From Fig. 2b  we can also observe the monotonous 
decrease in $R_{eq2}$ up to a field strength of $10\;a.u.$ 

Next let us discuss the existence of bound states. 
Herefore we have to investigate whether vibrational levels exist in the PECs discussed 
above. The determination of vibrational levels in the presence 
of a magnetic field is a complicated task since the 
Born-Oppenheimer approximation known from field free space 
breaks down in the presence of a magnetic field: The 
nuclear charges are treated as 'naked' charges which is an incomplete description since they 
are at least partially screened by the electrons against the magnetic field. 
In order to decribe this screening correctly, the diagonal terms of 
the nonadiabatic coupling elements have to be included in the nuclear equation of motion 
\cite{detmer:1995,schmelcher:1994,schmelcher2:1988}. 
The screening of the nuclear charges depends not only on the 
internuclear distance but also on the angle between the internuclear axis and 
the magnetic field. As a consequence the nuclear equation of motion 
is much more complex in the presence of a magnetic field 
compared to the field free space. In the present investigation we are dealing with the case 
of parallel internuclear and magnetic field axes and therefore cannot determine the 
exact vibrational levels. However we are able to provide estimations 
for the positions of the vibrational levels and on the basis of these estimates we can decide 
whether bound states exist with respect to the vibrational mode R. 

A lower estimate of the vibrational energy can be obtained by solving the nuclear 
equation of motion known from field free space $\frac{P^2}{2\mu} + V $,  
where the potential $V$ is given by the 
corresponding PEC in the magnetic field. The corresponding Schr\"odinger 
equation is then solved by using a discrete 
variable method \cite{colbert:1992}. An upper estimate for the energy of a vibrational state 
is obtained by simply adding the Landau energy of the nuclear equation of motion 
to the lower estimate of the energy level. 
Estimations concerning the existence of vibrational levels were performed for each PEC 
shown in all figures. The number of vibrational levels with quantum number 
$J = \Lambda = 0$ in the first well 
of the PEC of the $^1\Pi_g$ state  (see Fig. 2a) decreases monotonously with increasing field strength 
up to $B = 0.1\;a.u.$ Bound states exist in the entire range of field strengths 
$0.0  \leq B \lesssim 0.1\;a.u.$
For the second, i.e. outer, well bound states exist up to a field strength 
of $2\;a.u.$ In the range $5 \lesssim B \lesssim 50\;a.u.$ the lower estimate of the 
vibrational energy 
was found below the energy in the separated atom limit where the upper bound is above. 
Therefore, the existence of bound states depends 
on the detailed nuclear motion and cannot be decided within the present approach. 
For even larger magnetic fields ($B\gtrsim100\;a.u.$) no vibrational states exists in the PEC 
of the $^1\Pi_g$ state. 

\subsection{The $^3\Pi_g$ state}
 
The shape of the PEC of the $^3\Pi_g$ state  depending on the magnetic field strength is shown 
in Fig. \ref{fig3} and the corresponding data are given in Table \ref{tab2}. 
With increasing field strength the absolute value of the total energy 
in the dissociation limit monotonously increases 
due to the spin-Zeeman shift in the magnetic field. 
As in the case of the $^1\Pi_g$ state  we observe a decrease in the dissociation energy 
with increasing field strength from $E_d = 0.034554\;a.u.$ in field free space 
to $E_d = 0.014550\;a.u.$ for a field strength of $0.1\;a.u.$ For $B = 0.2$ and $0.5\;a.u.$ 
we encounter metastable states with respect to the dissociation into two 
hydrogen atoms. For these states, the difference of the total energy 
of the maximum and the minimum of the corrresponding PEC instead of the 
dissociation energy is given in Table \ref{tab2}. The position of the equilibrium internuclear 
distance is shifted to slightly larger values 
with increasing field strength, i.e. from $R_{eq} = 2.01\;a.u.$ in field free space to 
$2.15\;a.u.$ for $B = 0.5\;a.u.$ Simultaneously the position of the maximum $R_{max}$ 
decreases from $4.35\;a.u.$ in field free space to $2.76\;a.u.$ for $B = 0.5\;a.u.$ 
For field strengths $B \gtrsim 1\;a.u.$ no maxima or minima occur in the 
PEC of the $^3\Pi_g$ state . However, the PEC possesses two turning points for $B \geq 1\;a.u.$ 
which can be clearly seen in Fig. \ref{fig3} for $B = 1$ and $2\;a.u.$ 
The positions of these two turning points are shifted from $2.16$ and $2.98\;a.u.$ 
for $B = 1\;a.u.$ to $0.54$ and $0.8\;a.u.$ for $B = 100\;a.u.$ 
In contrast to the PECs of the $^1\Pi_g$ state , well pronounced turning points exist for 
arbitrary field strengths in the range $1 \lesssim B \lesssim 100\;a.u.$ 

The above considerations indicate that the singlet-triplet energy splitting between 
the $^1\Pi_g$ state  and $^3\Pi_g\;$ state for $M_s = 0$ depends on the magnetic field strength. 
To investigate this let us first consider small internuclear distances, 
i.e. the regime, for which the triplet state 
is lower in energy than the singlet. From Tables \ref{tab1} and \ref{tab2} we can see that 
the singlet-triplet energy splitting does not vary strongly for $0 \leq B \leq 0.05\;a.u.$ which 
is also confirmed by the equality of the equilibrium internuclear distances 
$R_{eq1} \left( ^1\Pi_g \right) = R_{eq} \left( ^3\Pi_g\right) $ 
for that range of field strengths. The equilibrium internuclear distances $R_{eq1}$ and $R_{eq}$ 
differ for $B \gtrsim 0.1\;a.u.$ where $R_{eq}$ is located at smaller internuclear distances. 
Figs. 2a and \ref{fig3} reveal an obvious difference for $B = 0.5\;a.u.$ 
where a metastable state occurs in the PEC of the $^3\Pi_g$ state  which has no 
counterpart in the corresponding PEC of the $^1\Pi_g$ state. 
A careful look at Fig. \ref{fig4} reveals that for $B \lesssim 0.2\;a.u.$ the singlet-triplet splitting 
differs not too much from the splitting in field free space and both PECs 
are therefore similar. 
With further increasing field strength the maximum of the energy splitting is shifted 
to smaller internuclear distances and becomes more and more pronounced. 
For a broad range of internuclear distances $R_c < R < \infty$ 
the singlet state is lower in energy than the triplet state and the position of the minimum 
and turnover in the singlet-triplet splitting decreases monotonously with increasing field strength. 
The maximum absolute value of the splitting shown in Fig. \ref{fig4} 
occurs at a field strength of $20\;a.u.$
 
Vibrational bound states for the PECs of the $^3\Pi_g$ state   were found to exist for $0 \leq B \leq 0.1\;a.u.$ 
As for the PEC of the $^1\Pi_g$ state  
the number of levels decreases with increasing field strength. 

\subsection{The $^1\Pi_u$ state}
 
The PEC of the $^1\Pi_u$ state  exhibits a strongly pronounced potential well in field free 
space whose minimum is located at $1.95\;a.u.$,
and a maximum for large internuclear distances, i.e. at $R = 9.03\;a.u.$ 
For the entire range of field strengths $0 \leq B \leq 100\;a.u.$ the minimum 
and maximum persist in the PEC of the $^1\Pi_u$ state . The corresponding data, i.e. total and dissociation energies
and equilibrium internuclear distances, are given in 
Table \ref{tab3} and the PEC is illustrated for different field strengths in Fig. \ref{fig5}. 
With increasing field strength the dissociation energy $E_d$ first 
decreases slightly for small values of $B$, i.e. from $E_d = 0.093220\;a.u.$ 
in field free space to $0.087080\;a.u.$ for $B = 0.5\;a.u.$ 
Simultaneously the position of the equilibrium internuclear distance $R_{eq}$ is shifted 
from $1.95$ to $1.80\;a.u.$ However, the dissociation energy increases strongly with 
further increasing field strength. At the same time the position of $R_{eq}$ is shifted to monotonously 
decreasing internuclear distances. As an example we mention the PEC of the $^1\Pi_u$ state  
for $B = 100\;a.u.$ where the dissociation energy amounts to $1.208626\;a.u.$ and the 
equilibrium internuclear distance is $0.40\;a.u.$ The PEC for of the $^1\Pi_u$ state  for $B = 100\;a.u.$ 
is also shown in Fig. \ref{fig5}. In this figure we particularly observe the strongly 
increasing dissociation energy for large values of $B$. 
For the position of the maximum $R_{max}$ in the PEC we obtain a monotonous behavior 
in the range $0 \leq B \leq 100\;a.u.$ First, $R_{max}$ exhibits only a minor decrease 
from $9.03$ to $8.29\;a.u.$ for $B = 0$ and $0.1\;a.u.$, respectively. 
Subsequently the position of $R_{max}$ decreases more rapidly with further increasing field strength, 
i.e in the range $0.1 \leq B \leq 100\;a.u.$ where $R_{max}$ is shifted from $8.29$ 
to $2.43\;a.u.$
Since the height of the maximum is very small, it cannot be observed in Fig. \ref{fig5}. 

Many vibrational states exist in the entire range of field strengths $0 \leq B \leq 100\;a.u.$ 
Compared to the number of levels in field free space the number of levels slightly 
increases with increasing magnetic field strength. 

\subsection{The $^3\Pi_u$ state}

The PEC of the $^3\Pi_u$ state  exhibits a deep potential well in field free space which is located at 
an internuclear distance of $1.96\;a.u.$ In the presence of a magnetic field 
the equilibrium internuclear distance $R_{eq}$ remains approximately constant for $0 \leq B \leq 0.01\;a.u.$
At the same time the dissociation energy $E_d$ varies only slightly, i.e. increases
from $0.115522$ to $0.112559\;a.u.$ for $B = 0$ and $B = 0.01\;a.u.$, respectively. 
For larger field strengths we observe a strong increase in the dissociation 
energy and a simultaneous decrease in the corresponding equilibrium internuclear distance. 
The corresponding data for the $^3\Pi_u$ state  are presented in Table \ref{tab4} and the PEC is shown 
for different field strengths in Fig. \ref{fig6}. The 
dissociation energy for $B = 100\;a.u.$ amounts to $1.811759\;a.u.$ (cf. Table \ref{tab4}) 
which documents the enormous increase in $E_d$ to be clearly seen also in Fig. \ref{fig6}. 
An interesting phenomenon can be observed around a field strength of $B = 100\;a.u.$ 
For that field strength regime the PEC of the $^3\Pi_u$ state  exibits a shallow hump with a 
maximum located at 
$R_{max} = 3.11\;a.u.$ and a second minimum at $R_{eq2} = 4.51\;a.u.$ 
(see Table \ref{tab4}). However, this additional minimum is very shallow and the dissociation 
energy amounts to only $4.604\times10^{-5}\;a.u.$ Therefore, both the second minimum 
and the maximum cannot be seen in Fig. \ref{fig6}. 

In the high field regime, i.e. for field strengths larger than $1\times10^{7}$ T, 
the $^3\Pi_u$ state  has been investigated by Ortiz and coworkers \cite{ortiz:1995} and Lai \cite{lai:1992}. 
The equilibrium internuclear distance $R_{eq}$ for the $^3\Pi_u$ state  at a field strength of 
$1\times10^{7}$ T was determined to be $0.51\;a.u.$ with a total ground state energy 
of $163.03\;eV$ at the equilibrium internuclear distance $R_{eq}$ \cite{ortiz:1995}. 
In the present investigation we performed a calculation for the same field strength 
$1\times10^{7}\;T\;\left( 42.54414\;a.u.\right) $ 
and obtained a slightly different equilibrium internuclear distance of $0.50\;a.u.$ 
with a somewhat lower total energy of $163.54409\;eV$. The difference in the total energy 
at $R = 0.50$ and $0.51\;a.u.$ within our calculations amounts to only $3.57\times10^{-4}\;eV$. 
The total energy in the separated atom limit within our calculations was determined 
to be $-4.798851\;a.u.$ 
Compared with the best available data in the literature (see Ref. \cite{ortiz:1995}) our result 
shows an improvement of approximately $0.31\%$ in the total energy and of $1.19\%$ in the 
corresponding dissociation energy. 

The number of vibrational levels for the first well, which is about $20$ in 
field free space, remains of the same order of magnitude for arbitrary field strengths 
$0 \leq B \leq 100\;a.u.$ For the second minimum occuring at a field strength of $100\;a.u.$, 
the lower estimate of the vibrational energy is located inside the well  
while the upper estimate lies above and we therefore cannot decide whether it 
accomodates a vibrationally bound state. 

\section{The ground state of the hydrogen molecule in a magnetic field} \label{ground}

The ground state of the hydrogen molecule in field free space is the $^1\Sigma_g^+$ 
singlet state. In the presence of a magnetic field, the diamagnetic term in the Hamiltonian 
(\ref{form1}) causes an increase in the total energy with increasing field strength. 
At the same time, due to the interaction of the total electronic spin with the magnetic field, 
the spin-Zeeman shift occuring for triplet states with $\left( M_s = -1 \right) $ 
lowers the total energy. As a result the 
$^1\Sigma_g$ singlet state is not expected to remain the global ground state of the 
parallel configuration for sufficiently strong fields. In Ref. \cite{detmer1:1997} the crossing 
between the PECs of the singlet $^1\Sigma_g$ and triplet $^3\Sigma_u$ state was determined to occur 
between $B = 0.1$ and $0.2\;a.u.$ For $B \gtrsim 0.2\;a.u.$ the PEC of the $^3\Sigma_u$ state 
is lower in energy than that of the $^1\Sigma_g$ state and therefore represents the 
global ground state of the $H_2$ molecule for a certain range of field strengths (see below). 
Since the PEC of the $^3\Sigma_u$ state is, apart from a very shallow van der Waals minimum, 
a purely repulsive curve the hydrogen molecule is unstable (unbound) in the corresponding 
regime of field strengths. For much higher field strengths ($B \gtrsim 3\times10^3\;a.u.$) 
it has been shown \cite{ortiz:1995} that the ground state of the parallel configuration 
is the $^3\Pi_u$ state. The field strength belonging to the $^3\Sigma_u\,/\,^3\Pi_u$ 
crossing is however not yet known. 

In the following we determine and discuss the crossings among the above-mentioned 
three states as a function of the field strength. In Figs. 7a and 7b 
we show the total energies at the corresponding equilibrium internuclear distances 
of the $^1\Sigma_g$ and $^3\Pi_u$ state as well as the 
total energy in the dissociation limit of the $^3\Sigma_u$ state as a function of the field strength. 
First we focus on the 
crossings of the $^1\Sigma_g$ with the $^3\Sigma_u$ and $^3\Pi_u$ states which is 
illustrated in Fig. 7a. The transition between the 
$^1\Sigma_g$ and $^3\Sigma_u$ state occurs at a field strength of approximately $0.18\;a.u.$
and the crossing field strength 
between the $^3\Sigma_u$ and $^3\Pi_u$ state is determined 
to be at $B \approx 0.39\;a.u.$ 
With further increasing magnetic field strength the total 
energy of the $^1\Sigma_g$ state increases strongly and in particular 
the total energy of the $^3\Pi_u$ state 
decreases more rapidly than that of the $^3\Sigma_u$ state. Therefore, a transition occurs 
between the two latter states which is illustrated in Fig. 7b. 
As can be seen from Fig. 7b, the transition field strength occurs to be at $B \approx 12.3\;a.u.$ 

In conclusion we encounter the following situation for the $H_2$ molecule oriented 
parallel to a magnetic field: For $B \lesssim 0.18\;a.u.$ the ground state is the strongly 
bound $^1\Sigma_g$ state. For an intermediate range of field strengths, i.e. 
for $0.18\lesssim B \lesssim 12.3\;a.u.$, the ground state is the unbound $^3\Sigma_u$ state. 
This state exhibits only a very shallow minimum 
which does not provide any vibrational level \cite{detmer1:1997}. Therefore the ground state  
of the parallel configuration is not bound in the regime $0.18\lesssim B \lesssim 12.3\;a.u.$ 
This result is of great importance for astrophysics in order to decide whether 
hydrogen molecules exist in the vicinity of white dwarfs. For magnetic fields 
$B \gtrsim 12.3\;a.u.$ the ground state is again a strongly bound, namely the $^3\Pi_u$ state 
and molecular hydrogen may exist in the vicinity of astrophysical objects which 
exhibit such huge magnetic field strengths.  

\section{Summary and conclusions} \label{summary}

In the present investigation we studied the electronic structure of the hydrogen molecule 
subjected to an external magnetic field which is 
oriented parallel to the internuclear axis. We hereby focused on the lowest electronic states of the $\Pi$ 
subspace, i.e. the lowest states with molecular magnetic quantum number equal to one. 
We discussed our basis set of nonorthogonal nonspherical Gaussian orbitals and briefly 
described the theoretical aspects of our CI calculations. 

First of all we discussed the general properties of the lowest $\Pi$ states in field 
free space and compared our results with the existing data in the literature. 
Our results show an overall relative accuracy better than  
$\lesssim 10^{-4}$ at the equilibrium internuclear distances 
of the PECs and even further improves with increasing internuclear distance. 
Compared to the results obtained for the lowest $\Sigma$ states in field free space \cite{detmer1:1997}, 
we approximately improved 
the accuracy in the energy by one order of magnitude. 
This is most probably due to the smaller portion of correlation energy 
for the $\Pi$ states. 

Next we considered the lowest $\Pi$ states in the presence of a magnetic field. 
For the $^1\Pi_g$ state  we observed a monotonous decrease in the dissociation energy with increasing 
field strength up to $0.1\;a.u.$ and a simultaneous minor increase 
in the equilibrium internuclear distance. 
At the same time the position of the occuring maximum was shifted to smaller 
values of the internuclear 
distance. For a field strength of $0.2\;a.u.$ a metastable state with 
respect to the dissociation into two hydrogen atoms exists in the PEC 
and for $B \gtrsim 0.5\;a.u.$ both the minimum 
and the maximum disappear. For the second, i.e. outer minimum  
the dissociation energy first increases up to $B \lesssim 1\;a.u$ and than decreases 
with further increasing field strength. 
In order to decide whether vibrationally bound states exist 
in the discussed PECs we determined upper and lower estimates for the vibrational energies. 
Up to a field strength of $0.1\;a.u.$ the $^1\Pi_g$ state was shown to accomodate a few vibrational levels. 
For the second well vibrationally bound states exist in the range $0 \leq B \lesssim 2\;a.u.$ 
For $5 \lesssim B \lesssim 50\;a.u.$ the existence of bound states depends on the 
detailed nuclear dynamics of the molecule. 
For even larger field strengths no vibrational levels were found. 

For the PEC of the $^3\Pi_g$ state  we observe a decrease in the dissociation energy and a simultaneous 
increase in the equilibrium internuclear distance with increasing field strength in the 
regime $0 \leq B \lesssim 0.1\;a.u.$ Metastable states with respect to the dissociation 
into two H atoms were shown to exist for the range $0.2 \lesssim B \lesssim 0.5\;a.u.$ 
For the position of the maximum in the PEC we observed a monotonous decrease with 
increasing field strength. For $B \gtrsim 1\;a.u.$ the maximum and the minimum disappear. 
We also studied the singlet-triplet energy splitting between the $^1\Pi_g$ and $^3\Pi_g\left( M_s = 0 \right) $
state as a function of the magnetic field strength. This is motivated by the fact that the corresponding 
PECs in field free space are almost indistinguishable for sufficiently small internuclear distances. 
We hereby focused on the rough shape of the PECs and neglected the properties 
due to the very shallow second minimum occuring in the PEC of the $^1\Pi_g$ state. 
In the presence of a magnetic field and with increasing field strength 
we found an increase of the splitting between these two states 
for small internuclear distances, i.e. for the region, where the corresponding PECs 
are almost identical in field free space. 
For larger internuclear distances we found a significant difference 
between the PECs of the $^1\Pi_g$ and $^3\Pi_g \left( M_s = 0 \right) $ state for 
$1 \lesssim B \lesssim 20\;a.u.$ For this regime of field strengths the PEC of the $^3\Pi_g$ state exhibits 
no extrema but two turning points which do not exist in the corresponding PEC of the 
$^1\Pi_g$ state. The maximum of the absolute value of the splitting was found to occur 
at a field strength of approximately $20\;a.u.$ 
Vibrationally bound states were shown to exist up to $B = 0.1\;a.u.$

For the PEC of the $^1\Pi_u$ state we first observe a decrease in the dissociation energy up to 
a field strength of $B \approx 0.5\;a.u.$ with increasing field strength 
and subsequently a strong increase in the dissociation 
energy with further increasing field strength. 
At the same time the equilibrium internuclear distance decreases monotonously. 
For the number of vibrational levels we found a minor increase by a few levels compared 
to the number of levels in field free space. 

Next we studied the PEC of the $^3\Pi_u$ state in the presence of a magnetic field. 
The dissociation energy and the equilibrium internuclear distance remain nearly 
constant for small values of B, i.e. for $0 \leq B \lesssim 0.1\;a.u.$ 
With further increasing field strength we observe a strong increase in the dissociation energy. 
An interesting phenomenon is the existence of an additional minimum and a maximum for a 
field strength regime around  $100\;a.u.$ Despite the shallowness of the corresponding well 
it may accomodate a vibrationally bound state. 
For the first, i.e. well-pronounced well vibrational states exist for 
all field strengths $0 \leq B \leq 100\;a.u.$ 

Finally we investigated the important transitions occuring among the 
$^1\Sigma_g$, $^3\Sigma_u$ and $^3\Pi_u$ states. 
It was shown recently \cite{detmer1:1997} that the strongly bound $^1\Sigma_g$ state 
is the global ground state of the 
hydrogen molecule for $B \lesssim 0.18\;a.u.$ 
and that the global ground state beyond $0.18\;a.u.$ is the unbound $^3\Sigma_u$ state 
up to some unknown critical field strength $B_c$. 
The $^3\Sigma_u$ state exhibits a purely repulsive PEC apart from a  
shallow van der Waal minimum which does not provide any vibrational level. 
An important result of the present investigation is that at approximately $12.3\;a.u.$ 
the ground state changes from the $^3\Sigma_u$ state to the strongly bound $^3\Pi_u$ state. 
This is of great importance 
for the chemistry in the atmosphere of certain degenerate astrophysical objects: 
it may help to decide whether molecular hydrogen can exist in the vicinity 
of white dwarfs. We emphasize that the present investigation was performed for the case 
of parallel internuclear and magnetic field axes. In order to draw a definite conclusion 
concerning the ground state of the $H_2$ molecule the electronic structure has to be 
investigated for arbitrary angles between the molecule and the magnetic field. 

The above results may also give rise to an experimental scenario. 
Let us consider configurations with an angle in the range $0 < \theta < 90^o$, 
i.e. the internuclear axis is inclined with respect to the magnetic field. 
Since the electronic potential energy depends on both the internuclear distance $R$ as well as 
the angle $\theta$ we are dealing with two dimensional potential energy surfaces. 
The only remaining symmetry for such configurations for a homonuclear diatomic molecule 
is the parity, i.e. the corresponding molecular symmetry 
group is $C_i$. We now briefly address the question whether different electronic states may interact 
strongly through the nuclear motion. This problem was investigated in some detail in 
Ref. \cite{schmelcher:1990}. It was shown that 
for a homonuclear diatomic molecule a two dimensional avoided crossing occurs for 
electronic states with the same spatial symmetry and in particular 
that {\sl conical intersections} can occur 
at $\theta = 0$ or $90^o$ for crossings of states with the same parity. 

For $\theta \neq 0^o$ 
we expect a crossing without interaction between the lowest state with gerade and 
the lowest state with ungerade parity. 
More interestingly we expect a conical intersection to occur at $\theta = 0^o$ 
for the PECs of the two lowest states $\left( ^3\Pi_u - ^3\Sigma_u \right) $ with ungerade parity. 
The hydrogen molecule in the presence of a magnetic field therefore is an example for a simple 
system for which the {\sl ground state} in strong fields exhibits a {\sl conical intersection}. 
From an experimental point of view we are hereby not restricted to the case of static external 
magnetic fields but could also use pulsed fields. 
The time scale of the pulse is usually orders of magnitude larger than 
the time scale of the nuclear motion near a conical intersection. For the case of the 
hydrogen molecule, the field strength at which a conical intersection 
of the lowest states with ungerade parity may be observed, 
is $B \gtrsim 0.1\;a.u.$ which corresponds to a few ten thousand Tesla. 
Pulsed magnetic fields of that strength may be reachable in the laboratory 
in the nearer future. 

\section{Acknowledgements}

The Deutsche Forschungsgemeinschaft is gratefully acknowledged for financial support. 
One of the authors (P.S.) acknowledges many fruitful discussions during the CECAM workshop 
'Atoms in strong magnetic fields' in Lyon. 
Computer time has been generously provided by the Rechenzentrum Karlsruhe 
and the Rechenzentrum Heidelberg.

\begin{figure}
\caption{(a) {PECs for $B = 0.0\,,\,0.1\,,\,0.2\,,\,0.5\,,\,1.0\,,\,10.0$ and $100.0\;a.u.$ 
of the lowest $^1\Pi_g$ state illustrating the first minimum and the maximum. The energy is 
given with respect to the dissociation limit, i.e. 
$E\left(R\right) = E_t\left(R\right) - \lim\limits_{R\to\infty} E_t\left(R\right)$ \newline
(b) PECs for $B = 0.0\,,\,0.2\,,\,1.0\,,\,10.0$ and $100.0\;a.u.$ 
for the lowest $^1\Pi_g$ state illustrating the second minimum. The energy is 
shown with respect to the dissociation limit, i.e. 
$E\left(R\right) = E_t\left(R\right) - \lim\limits_{R\to\infty} E_t\left(R\right)$}}
\label{fig2}
\end{figure}

\begin{figure}
\caption{PECs for $B = 0.0\,,\,0.1\,,\,0.2\,,\,0.5\,,\,1.0\,,\,2.0$ and $100.0\;a.u.$ 
for the lowest $^3\Pi_g$ state. The energy is 
shown with respect to the dissociation limit, i.e. 
$E\left(R\right) = E_t\left(R\right) - \lim\limits_{R\to\infty} E_t\left(R\right)$}
\label{fig3}
\end{figure}

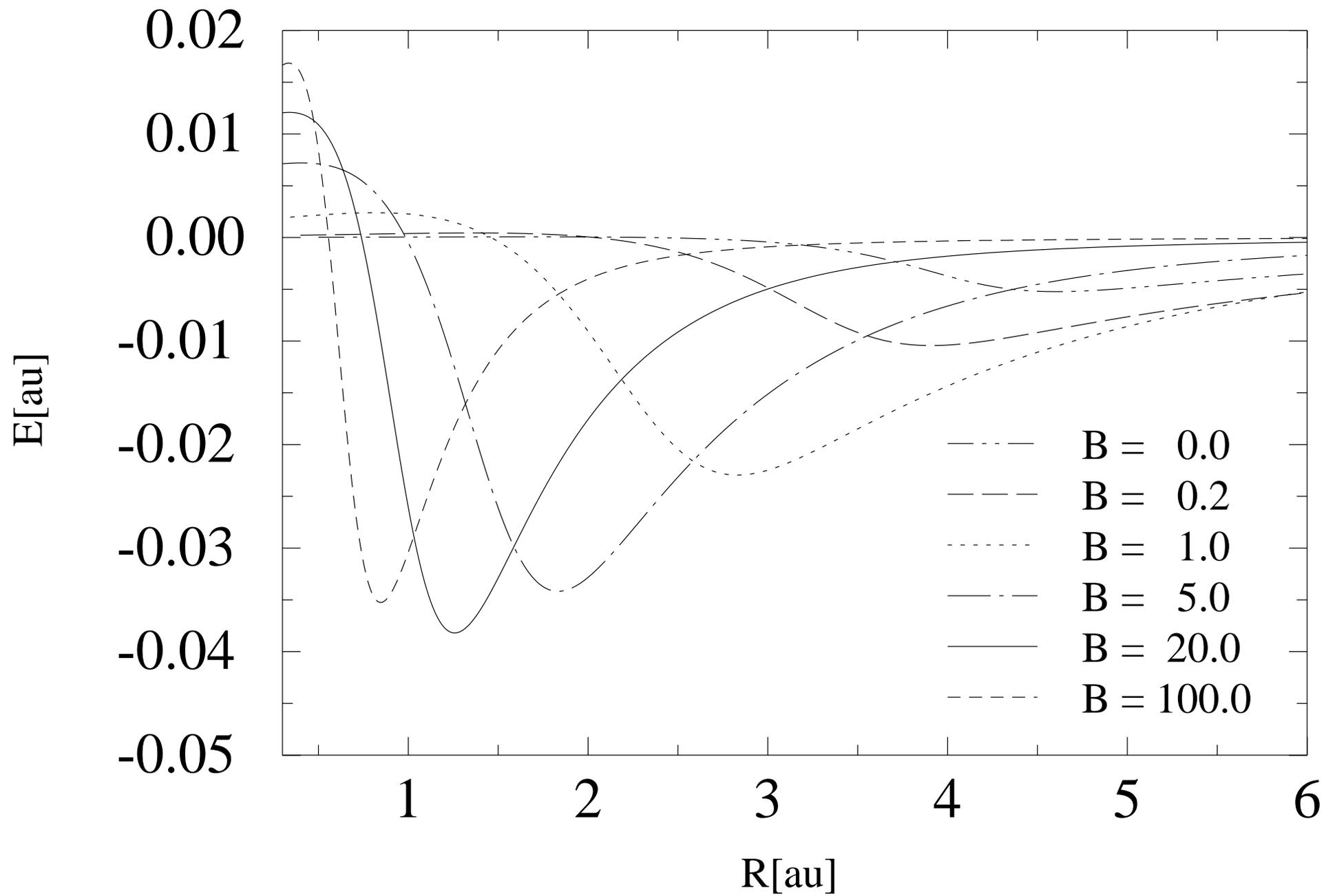
\begin{figure}
\caption{Singlet-triplet splitting between the $^1\Pi_g$ and $^3\Pi_g\left(M_s = 0\right)$ state for 
$B = 0.0\,,\,0.2\,,\,1.0\,,\,5.0$ $20.0$ and $100.0\;a.u.$}
\label{fig4}
\end{figure}

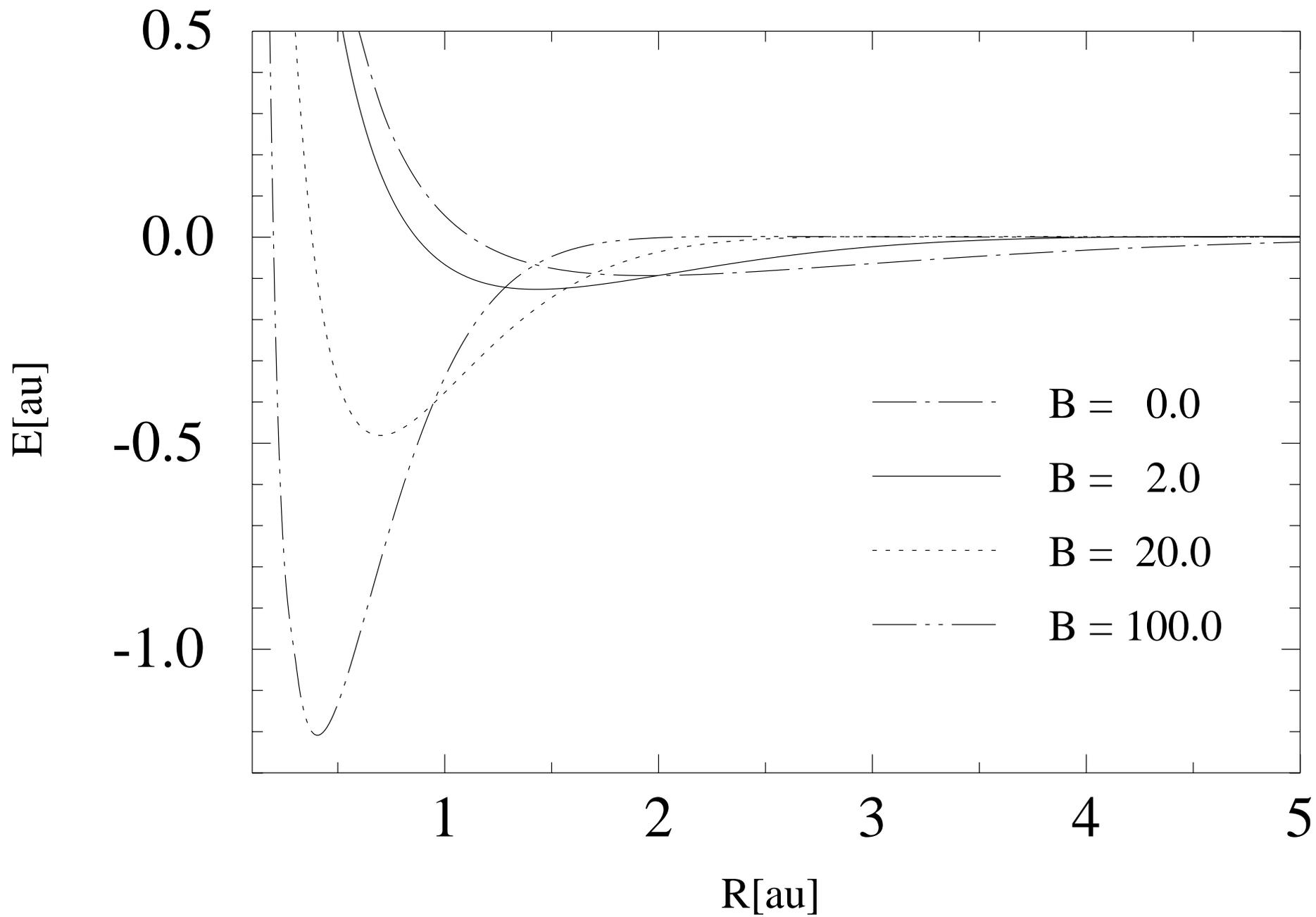
\begin{figure}
\caption{PECs for $B = 0.0\,,\,2.0\,,\,20.0$ and $100.0\;a.u.$ 
for the lowest $^1\Pi_u$ state. The energy is 
shown with respect to the dissociation limit, i.e. 
$E\left(R\right) = E_t\left(R\right) - \lim\limits_{R\to\infty} E_t\left(R\right)$}
\label{fig5}
\end{figure}

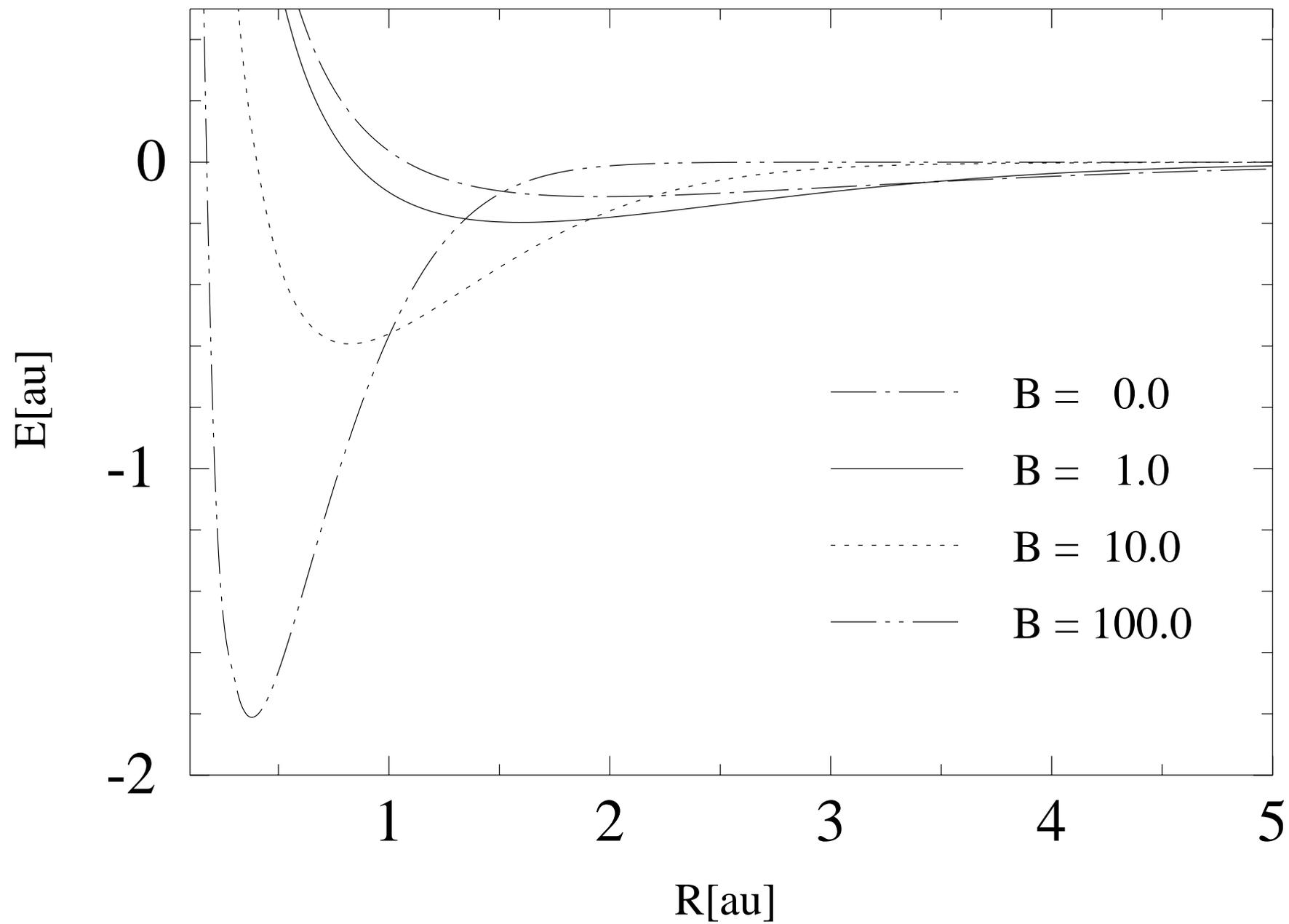
\begin{figure}
\caption{PECs for $B = 0.0\,,\,1.0\,,\,10.0$ and $100.0\;a.u.$ 
for the lowest $^3\Pi_u$ state. The energy is 
shown with respect to the dissociation limit, i.e. 
$E\left(R\right) = E_t\left(R\right) - \lim\limits_{R\to\infty} E_t\left(R\right)$}
\label{fig6}
\end{figure}

\begin{figure}
\caption{(a) Transitions between the lowest $^1\Sigma_g$ and the $^3\Sigma_u$ and $^3\Pi_u$ states  
showing the total energy of the $^1\Sigma_g$ and $^3\Pi_u$ states (bound) 
at the corresponding equilibrium internuclear distance and the total energy in the dissociation limit 
of the $^3\Sigma_u$ state (unbound) \newline
         (b) Transition between the lowest $^3\Sigma_u$ and $^3\Pi_u$ state 
showing the total energy in the dissociation limit of the $^3\Sigma_u$ state (unbound) 
and the total energy at the corresponding equilibrium internuclear distance of the $^3\Pi_u$ state 
(bound)}
\label{fig7}
\end{figure}

\begin{table}
\caption{Data for the lowest $^1\Pi_g$ state: Total energies $E_{t1},E_{t2}$ 
and dissociation energies $E_{d1},E_{d2}$ 
at the equilibrium internuclear distance, 
the equilibrium internuclear distances $R_{eq1},R_{eq2}$,
the positions $R_{max}$ and total energies $E_{max}$ of the maxima  
and the total energies in the dissociation limit $\lim\limits_{R\to\infty} E_{t}$ 
as a function of the field strength $0 \leq B \leq 100$ (all quantities are given in atomic units).}
{\squeezetable
\begin{tabular}{dddddddddd}
\multicolumn{1}{c}{\rule[-5mm]{0mm}{11mm}{\raisebox{-0.5ex}[0.5ex]{B}}} & 
\multicolumn{1}{c}{\raisebox{-0.5ex}[0.5ex]{$R_{eq1}$}} &
\multicolumn{1}{c}{\raisebox{-0.5ex}[0.5ex]{$E_{d1}$}} &
\multicolumn{1}{c}{\raisebox{-0.5ex}[0.5ex]{$E_{t1}$}} &
\multicolumn{1}{c}{\raisebox{-0.5ex}[0.5ex]{$R_{eq2}$}} &
\multicolumn{1}{c}{\raisebox{-0.5ex}[0.5ex]{$E_{d2}$}} &
\multicolumn{1}{c}{\raisebox{-0.5ex}[0.5ex]{$E_{t2}$}} &
\multicolumn{1}{c}{\raisebox{-0.5ex}[0.5ex]{$R_{max}$}} &
\multicolumn{1}{c}{\raisebox{-0.5ex}[0.5ex]{$E_{max}$}} &
\multicolumn{1}{c}{$\lim\limits_{R\to\infty} E_{tot}$} \\ 
0.0  & 2.01 & 0.034502 & $-$0.659501 & 8.14 & 8.038$\times10^{-4}$ & $-$0.625803 & 4.24 & $-$0.616547 & $-$0.624999 \\
0.001& 2.01 & 0.034501 & $-$0.659997 & 8.14 & 8.039$\times10^{-4}$ & $-$0.626230 & 4.24 & $-$0.617068 & $-$0.625496 \\
0.005& 2.01 & 0.034268 & $-$0.661686 & 8.14 & 8.045$\times10^{-4}$ & $-$0.628222 & 4.24 & $-$0.618976 & $-$0.627418 \\
0.01 & 2.01 & 0.033959 & $-$0.663634 & 8.14 & 8.087$\times10^{-4}$ & $-$0.630484 & 4.23 & $-$0.621186 & $-$0.629675 \\
0.05 & 2.02 & 0.026191 & $-$0.668625 & 7.95 & 9.100$\times10^{-4}$ & $-$0.643344 & 4.06 & $-$0.632934 & $-$0.642434 \\
0.1  & 2.04 & 0.014450 & $-$0.662821 & 7.60 & 1.132$\times10^{-3}$ & $-$0.649503 & 3.77 & $-$0.636987 & $-$0.648371 \\
0.2  & 2.09 & 0.011130\tablenote{This electronic state is metastable with respect to the dissociation into two H atoms. The difference between the first minimum and the maximum of the PEC is given instead of the dissociation energy.} & $-$0.636414 & 7.00 & 1.586$\times10^{-3}$ & $-$0.642506 & 3.28 & $-$0.625284 & $-$0.640920 \\
0.5  & &     &      & 5.99 & 2.437$\times10^{-3}$ & $-$0.549125 & &      & $-$0.546687 \\
1.0  & &     &      & 5.25 & 2.853$\times10^{-3}$ & $-$0.290617 & &      & $-$0.287765 \\
2.0  & &     &      & 4.66 & 2.702$\times10^{-3}$ &  0.375473   & &      &  0.378174 \\
5.0  & &     &      & 4.18 & 1.707$\times10^{-3}$ &  2.758062   & &      &  2.759769 \\
10.0 & &     &      & 4.05 & 8.791$\times10^{-4}$ &  7.125902   & &      &  7.126781 \\
20.0 & &     &      & 4.11 & 3.751$\times10^{-4}$ & 16.318719   & &      & 16.319094 \\
50.0 & &     &      & 4.32 & 1.137$\times10^{-4}$ & 44.925181   & &      & 44.925295 \\
100.0& &     &      & 4.55 & 4.57 $\times10^{-5}$ & 93.575395   & &      & 93.575441 \\
\end{tabular}
}
\label{tab1}
\end{table}

\begin{table}
\caption{Data for the lowest $^3\Pi_g$ state: Total energy $E_{t}$ 
and dissociation energy $E_{d}$ 
at the equilibrium internuclear distance, 
the equilibrium internuclear distances $R_{eq}$,
the positions $R_{max}$ and total energies $E_{max}$ of the maximum 
and the total energies in the dissociation limit $\lim\limits_{R\to\infty} E_{t}$ 
as a function of the field strength $0 \leq B \leq 100$ (all quantities are given in atomic units).}
\begin{tabular}{ddddddd}
\multicolumn{1}{c}{\rule[-5mm]{0mm}{11mm}{\raisebox{-0.5ex}[0.5ex]{B}}} & 
\multicolumn{1}{c}{\raisebox{-0.5ex}[0.5ex]{$R_{eq}$}} &
\multicolumn{1}{c}{\raisebox{-0.5ex}[0.5ex]{$E_{d}$}} &
\multicolumn{1}{c}{\raisebox{-0.5ex}[0.5ex]{$E_{t}$}} &
\multicolumn{1}{c}{\raisebox{-0.5ex}[0.5ex]{$R_{max}$}} &
\multicolumn{1}{c}{\raisebox{-0.5ex}[0.5ex]{$E_{max}$}} &
\multicolumn{1}{c}{$\lim\limits_{R\to\infty} E_{tot}$} \\ 
0.0  & 2.01 & 0.034554 & $-$0.659553 & 4.35 & $-$0.611691 & $-$0.624999 \\
0.001& 2.01 & 0.034551 & $-$0.661047 & 4.35 & $-$0.613211 & $-$0.626496 \\
0.005& 2.01 & 0.034320 & $-$0.666738 & 4.35 & $-$0.619112 & $-$0.632418 \\
0.01 & 2.01 & 0.034012 & $-$0.673687 & 4.34 & $-$0.626296 & $-$0.639675 \\
0.05 & 2.02 & 0.026271 & $-$0.718705 & 4.18 & $-$0.677419 & $-$0.692434 \\
0.1  & 2.03 & 0.014550 & $-$0.762921 & 3.93 & $-$0.730337 & $-$0.748371 \\
0.2  & 2.06 & 0.019590\tablenotemark[1] & $-$0.836343 & 3.53 & $-$0.816753 & $-$0.840920 \\ 
0.5  & 2.15 & 0.002483\tablenotemark[1] & $-$1.009346 & 2.76 & $-$1.006863 & $-$1.046687 \\
1.0  &      &          &             &      &             & $-$1.287765 \\
2.0  &      &          &             &      &             & $-$1.621825 \\
5.0  &      &          &             &      &             & $-$2.240231 \\
10.0 &      &          &             &      &             & $-$2.873218 \\
20.0 &      &          &             &      &             & $-$3.680905 \\
50.0 &      &          &             &      &             & $-$5.074705 \\
100.0&      &          &             &      &             & $-$6.424559 \\
\end{tabular}
\tablenotetext[1]{These electronic states are metastable with respect to the dissociation into two H atoms. The difference between the minimum and the maximum of the PEC is given instead of the dissociation energy.}
\label{tab2}
\end{table}

\begin{table}
\caption{Data for the lowest $^1\Pi_u$ state: Total energy $E_{t}$ 
and dissociation energy $E_{d}$ 
at the equilibrium internuclear distance, 
the equilibrium internuclear distances $R_{eq}$,
the positions $R_{max}$ and total energies $E_{max}$ of the maxima  
and the total energies in the dissociation limit $\lim\limits_{R\to\infty} E_{t}$ 
as a function of the field strength $0 \leq B \leq 100$ (all quantities are given in atomic units).}
\begin{tabular}{ddddddd}
\multicolumn{1}{c}{\rule[-5mm]{0mm}{11mm}{\raisebox{-0.5ex}[0.5ex]{B}}} & 
\multicolumn{1}{c}{\raisebox{-0.5ex}[0.5ex]{$R_{eq}$}} &
\multicolumn{1}{c}{\raisebox{-0.5ex}[0.5ex]{$E_{d}$}} &
\multicolumn{1}{c}{\raisebox{-0.5ex}[0.5ex]{$E_{t}$}} &
\multicolumn{1}{c}{\raisebox{-0.5ex}[0.5ex]{$R_{max}$}} &
\multicolumn{1}{c}{\raisebox{-0.5ex}[0.5ex]{$E_{max}$}} &
\multicolumn{1}{c}{$\lim\limits_{R\to\infty} E_{tot}$} \\ 
0.0  & 1.95 & 0.093220 & $-$0.718219 & 9.03 & $-$0.624528 & $-$0.624999 \\
0.001& 1.95 & 0.093215 & $-$0.718711 & 9.02 & $-$0.625025 & $-$0.625496 \\
0.005& 1.95 & 0.093178 & $-$0.720596 & 9.02 & $-$0.626945 & $-$0.627418 \\
0.01 & 1.95 & 0.093163 & $-$0.722838 & 9.01 & $-$0.629200 & $-$0.629675 \\
0.05 & 1.94 & 0.092348 & $-$0.734782 & 8.73 & $-$0.641874 & $-$0.642434 \\
0.1  & 1.93 & 0.090740 & $-$0.739111 & 8.29 & $-$0.647645 & $-$0.648371 \\
0.2  & 1.90 & 0.087969 & $-$0.728889 & 7.56 & $-$0.639842 & $-$0.640920 \\
0.5  & 1.80 & 0.087080 & $-$0.633767 & 6.47 & $-$0.544981 & $-$0.546687 \\
1.0  & 1.65 & 0.097448 & $-$0.385213 & 5.72 & $-$0.285750 & $-$0.287765 \\
2.0  & 1.41 & 0.126544 &  0.251630 & 5.05 &  0.380278 &  0.378174 \\
5.0  & 1.11 & 0.207874 &  2.551895 & 4.26 &  2.761807 &  2.759769 \\
10.0 & 0.89 & 0.315707 &  6.811074 & 3.73 &  7.128738 &  7.126781 \\
20.0 & 0.70 & 0.481139 & 15.837955 & 3.27 & 16.320973 & 16.319094 \\
50.0 & 0.51 & 0.822810 & 44.102485 & 2.75 & 44.927090 & 44.925295 \\
100.0& 0.40 & 1.208626 & 92.366815 & 2.43 & 93.577178 & 93.575441 \\
\end{tabular}
\label{tab3}
\end{table}

\begin{table}
\caption{Data for the lowest $^3\Pi_u$ state: Total energy $E_{t}$ and dissociation energy $E_{d}$
at the equilibrium internuclear distance, 
the equilibrium internuclear distances $R_{eq}$,
and the total energies in the dissociation limit $\lim\limits_{R\to\infty} E_{t}$ 
as a function of the field strength $0 \leq B \leq 100$ (all quantities are given in atomic units).}
\begin{tabular}{ddddd}
\multicolumn{1}{c}{\rule[-5mm]{0mm}{11mm}{\raisebox{-0.5ex}[0.5ex]{B}}} & 
\multicolumn{1}{c}{\raisebox{-0.5ex}[0.5ex]{$R_{eq}$}} &
\multicolumn{1}{c}{\raisebox{-0.5ex}[0.5ex]{$E_{d}$}} &
\multicolumn{1}{c}{\raisebox{-0.5ex}[0.5ex]{$E_{t}$}} &
\multicolumn{1}{c}{$\lim\limits_{R\to\infty} E_{tot}$} \\ 
  0.0   & 1.96 & 0.112522 & $-$0.737521 & $-$0.624999  \\
  0.001 & 1.96 & 0.112533 & $-$0.739029 & $-$0.626496  \\
  0.005 & 1.96 & 0.112544 & $-$0.744962 & $-$0.632418  \\
  0.01  & 1.96 & 0.112559 & $-$0.752234 & $-$0.639675  \\
  0.05  & 1.95 & 0.113862 & $-$0.806296 & $-$0.692434  \\
  0.1   & 1.95 & 0.116912 & $-$0.865283 & $-$0.748371  \\
  0.2   & 1.91 & 0.125028 & $-$0.965948 & $-$0.840920  \\
  0.5   & 1.79 & 0.153029 & $-$1.199716 & $-$1.046687  \\
  1.0   & 1.59 & 0.196553 & $-$1.484318 & $-$1.287765  \\
  2.0   & 1.35 & 0.267435 & $-$1.889260 & $-$1.621285  \\
  5.0   & 1.03 & 0.418679 & $-$2.658910 & $-$2.240231  \\
 10.0   & 0.82 & 0.593026 & $-$3.466244 & $-$2.873218  \\
 20.0   & 0.65 & 0.838135 & $-$4.519040 & $-$3.680905  \\
 50.0   & 0.48 & 1.308786 & $-$6.383491 & $-$5.074705  \\
100.0 \tablenotemark[1] & 0.38 & 1.811759 & $-$8.236318 & $-$6.424559  \\
\end{tabular}
\label{tab4}
\tablenotetext[1]{This electronic state exhibits a second minimum and an additional maximum \newline $R_{eq2} = 4.51\;a.u.$ ; $E_{d2} = 4.604\times10^{-5}$ ; $E_{t2} = -6.424605$ \newline $R_{max} = 3.11$ ; $E_{max} = -6.424531$}
\end{table}

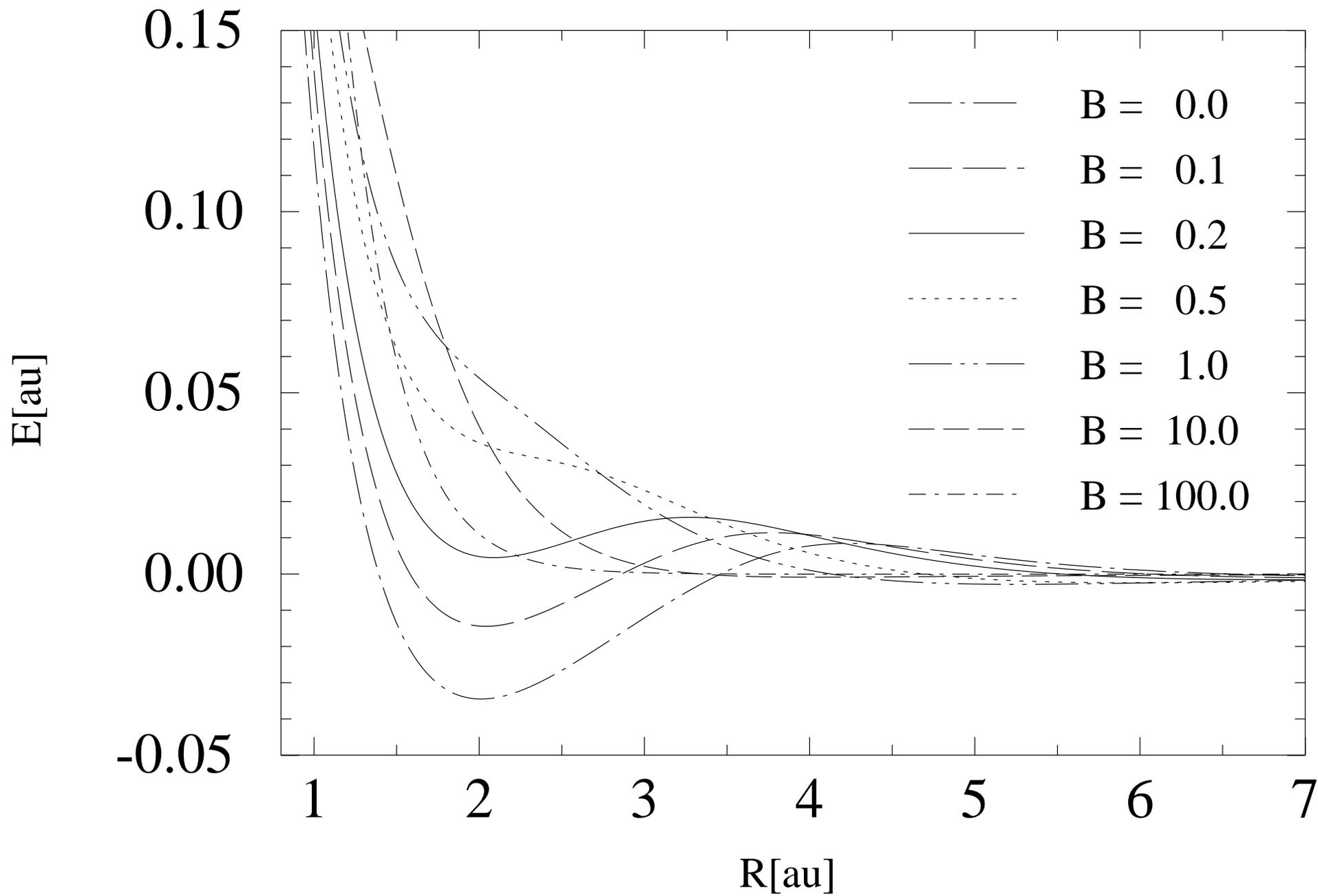
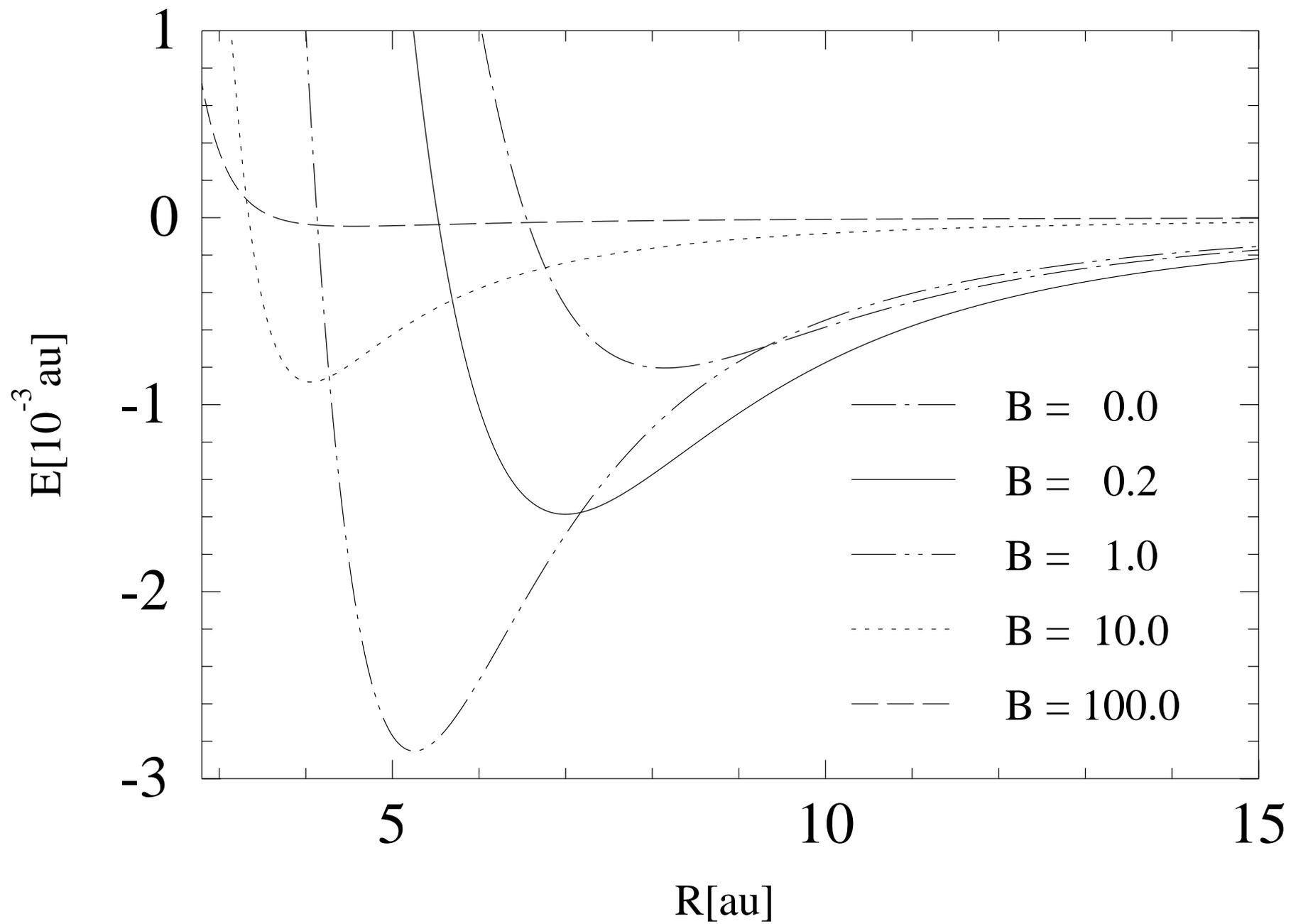
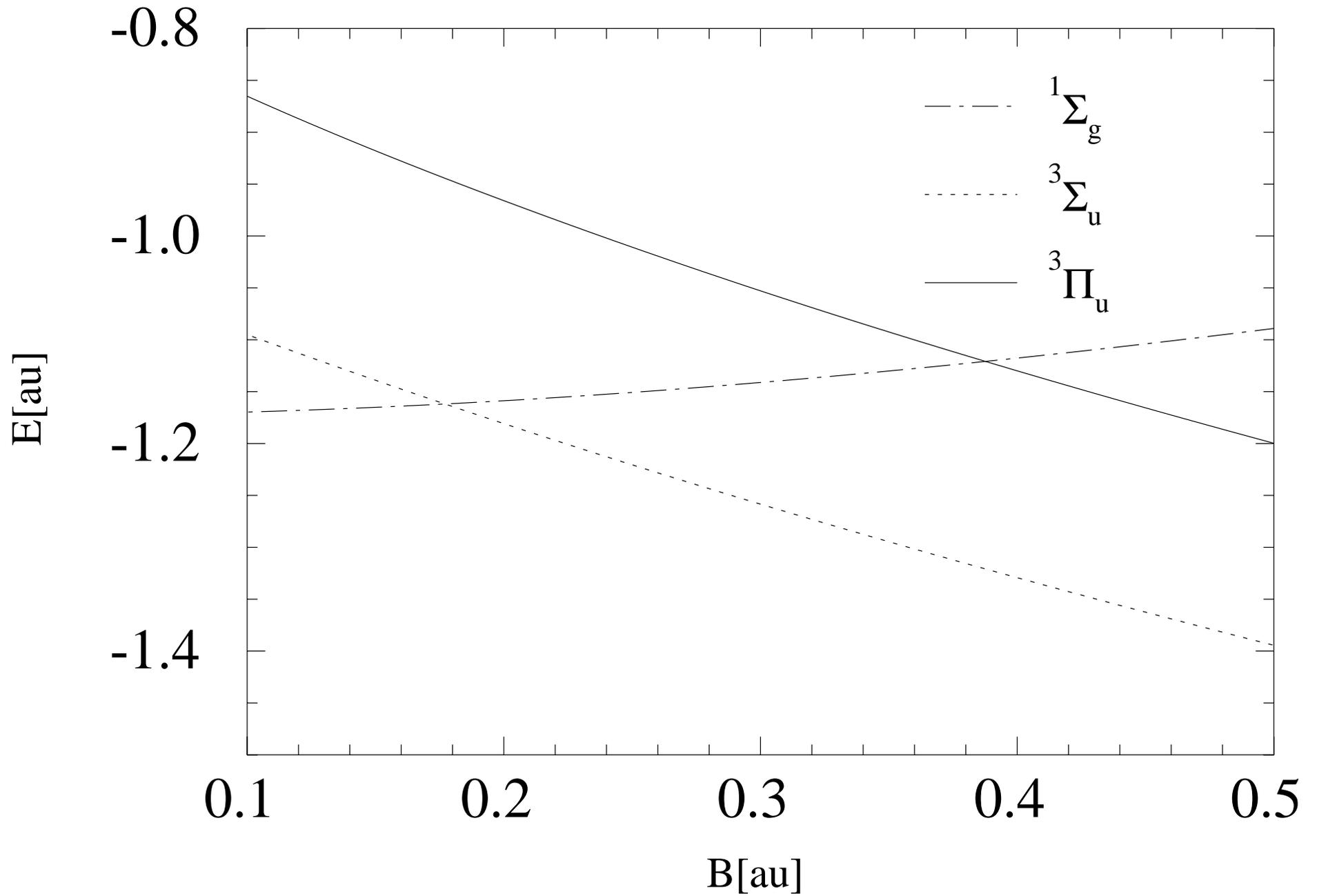
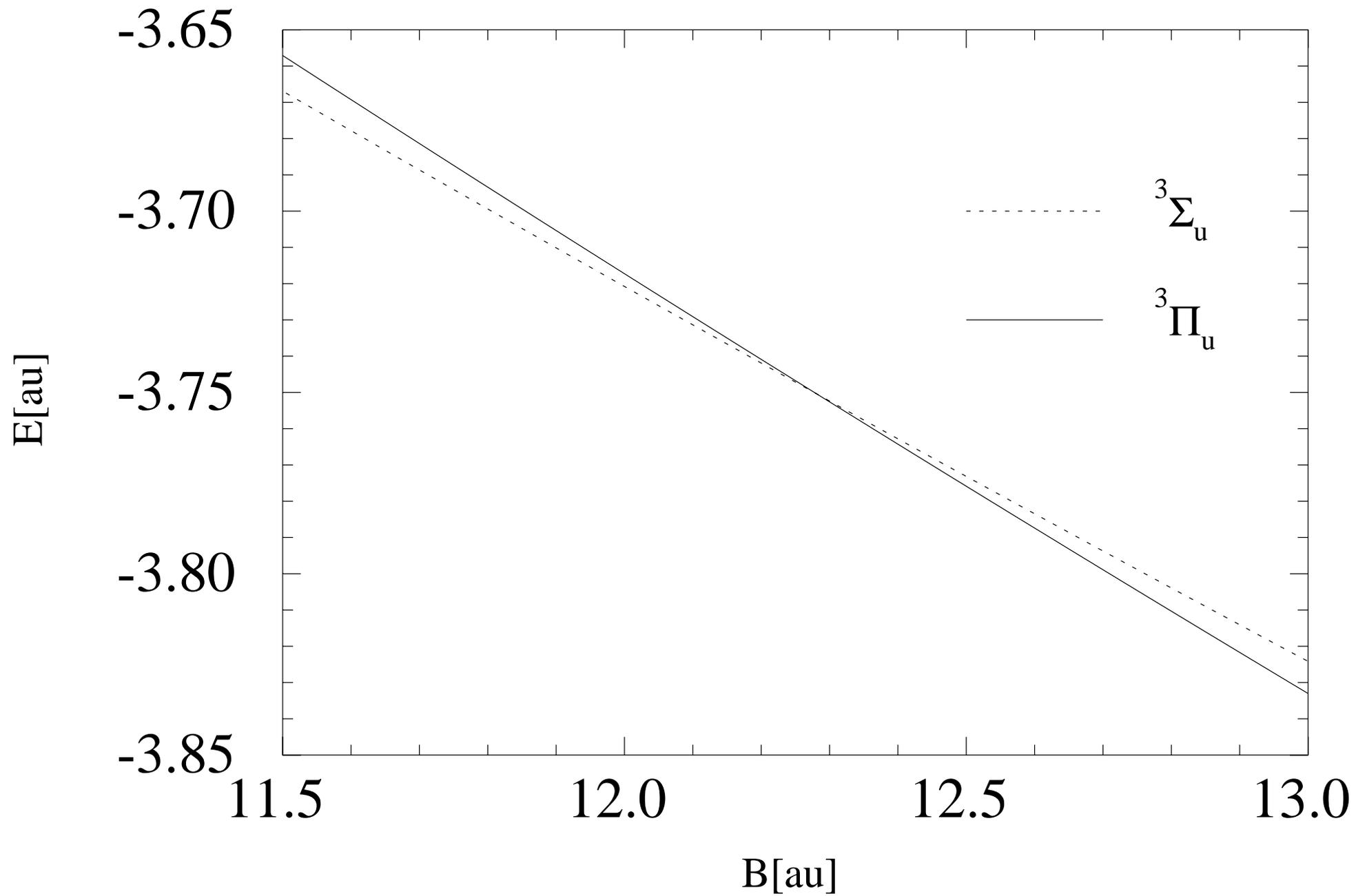
\end{document}